# Neural correlates of learned categorical perception of visual stimuli with local features

Pérez-Gay, F.[1,2], Sicotte, T.[2], Goulet, N[2]. Kang, X.[3],Harnad, S.[1,2,3]

[1]McGill University, [2]Université du Québec à Montréal, [3]University of Southampton

**Abstract:** Learning to categorize requires distinguishing category members from non-members by detecting the features that covary with membership. Whether this process can induce changes in perception is still a matter of debate. In prior studies, we reported Learned Categorical Perception (Learned CP) effects in the form of between-category separation and within-category compression in perceived similarity after training subjects to categorize visual stimuli with distributed, holistic features. These effects were correlated with changes in an early, perceptual component of Event Related Potentials (ERPs). Using the same methodology, in this experiment we trained 96 subjects to sort line drawings of fish with local, verbalizable features into two categories by trial and error with corrective feedback. We tested for Learned CP effects and their neural correlates by measuring subjects' pairwise dissimilarity judgments and (ERPs) before and after the training. With the same frequency of trials and feedback, about 40% of the participants succeeded in learning the categories ("learners") while the rest did not ("non-learners"). Learners showed a) significant between-category separation and within-category compression after training compared to before and b) an increase in a late parietal ERP positivity (LPC; 400-600 ms) and an early, occipital positivity (P1; 80-140 ms), correlated with categorization accuracy and degree of Learned CP. These behavioral and neural changes after learning a category, absent in Non Learners, provide further evidence that category learning can modulate perceptual processes, regardless of the nature of the visual features. We complement our experimental results with a neural net model using the same stimuli.

## 1. Introduction

The world we live in is full of inter-confusable things and situations. To survive, our brains have evolved the ability to detect patterns and identify shared features across objects and experiences (Mattson, 2014), allowing us to sort them into distinct, meaningful groups, towards which we can act in the same way. This process, through which we learn to "do the right thing with the right kind of thing", is called categorization (Harnad, 2005), and the result are "categories", the building blocks of human cognition.

Features are sensory properties of things - size, color, shape, texture, loudness or odor- or motor affordances -actions we can perform with those things (Gibson, 1979; Tucker & Ellis, 1998). Apart from those categories for which we are born with innate features-detectors (colors, phonemes or basic facial expressions), the majority of our categories are learned through experience. To learn a new category, our neural networks must detect the features that "co-vary" with category membership- and inhibit or ignore other features, which presence or absence is independent of category membership (Gao, Cai, Li, Zhang, & Li, 2016; Pérez-Gay, Sicotte, Thériault, & Harnad, 2019).

"Categorical perception" (CP) occurs when members of different categories appear more different from one another (separation) or members of the same category appear more alike (compression) (Goldstone & Hendrickson, 2009; Harnad, 2003) making differences in how things look or sound to us to depend on their category membership *rather than just on their physical structure alone.* This effect was initially described for colors, facial expressions, and phonemes, that do not depend on experience. Darwinian evolution has equipped us with inborn "feature-filters" for the information that reaches our senses that allow us to detect the features that covary with (and are thus predictive of) these categories. In turn, "Learned CP", occurs when between-category separation or within-category compression are induced by learning. The effect is detected by comparing perceived differences or discrimination indices between and within categories *before and after* having learned them (De Leeuw, Andrews, Livingston, & Chin, 2016).

In the last decade, functional neuroimaging has provided evidence of changes induced by visual category learning in brain areas associated with perceptual processing. Jiang et al., 2007 reported that learning to categorize a set of cars induced changes in lateral occipital cortex activity and Folstein, Palmeri, & Gauthier, 2013 reported changes in extrastriate occipital cortex activity when these car stimuli differed on the category-distinguishing feature. While these and other recent neuroimaging studies support these top-down effects of higher cognitive processes (including category learning) in earlier, perceptual processes (Ester, Sprague, Serences, & Ester, 2017.; Lupyan, 2012; Maier & Abdel Rahman, 2019; van der Linden, Wegman, & Fernández, 2014), the "true perceptual nature" of these effects continues to be questioned.

According to the "cognitive impenetrability" view, higher-order processes such as categorization could only influence pre-perceptual (attention-related) or post-perceptual (working memory or decisional) processes, but not perception itself (Firestone & Scholl, 2015; Gross, 2017; Pylyshyn, 1999; Raftopoulos, 2017). While the fMRI studies mentioned above give some idea of "where" learned CP may occur in the brain, their lack of temporal resolution makes it difficult to answer essential questions about the processing stage associated with these changes: are they truly perceptual, or do they arise from pre/post-perceptual cognitive processes – i.e., attention bias or thinking about things differently rather than seeing them differently? If they involve some sort of learned feature-filtering, what is the mechanism behind it and what are its neural substrates?

Because attentional and perceptual processes appear to operate on a scale of "tens of milliseconds" (Woodman, 2010), electrophysiological methods like Event Related Potentials (ERPs) are more suitable for this grain of temporal resolution than functional neuroimaging methods: Earlier perceptual effects can be distinguished from later decisional or higher-cognitive effects in ERPs. The shape, amplitude and latency of earlier ERP components are related to sensory and perceptual processes potentially relevant to learned CP such as selective attention (Hillyard & Anllo-Vento, 1998; Luck & Kappenman, 2012) visual discrimination (Fedota, Mcdonald, Roberts, & Parasuraman, 2012; Hopf, Voge, Woodman, Heinze, & Luck, 2002), feature selection (Ansorge, Kiss, Worschech, & Eimer, 2011; Han, Liu, Yund, & Woods, 2000;

Luck & Kappenman, 2012) and spatial cueing (Baumgartner, Graulty, Hillyard, & Pitts, 2018; He, Humphreys, Fan, Chen, & Han, 2008), allowing us to draw conclusions on the "true perceptual" nature of the changes induced by category learning.

Previous studies have shown changes induced by categorization in early ERPs (namely the P1 and the N1, considered as indices of early visual processing. Franklin and colleagues have shown an effect of color categories on the P1 using an oddball paradigm Forder, He, & Franklin, 2017; Holmes, Franklin, Clifford, & Davies, 2009; Thierry, Athanasopoulos, Wiggett, Dering, & Kuipers, 2009), and Maier, et. al (2019, 2014) have reported similar effect with images of rare objects, grouped into categories by either arbitrary verbal labels or semantic descriptions. In turn, other studies of category learning have showed changes in the occipital N1 -throughout a trial and error category learning task (Curran, Tanaka, & Weiskopf, 2002; Morrison, Reber, Bharani, & Paller, 2015; Pérez-Gay Juárez, et al., 2019). However, only Maier (2019) and Pérez-Gay Juárez (2019) linked these electrophysiological effects with behavioral Learned CP effects, corroborating that the size of CP effects correlates with the size of early physiological components (P1 and N1, respectively).

### 1.1 The present study

In our prior studies (Pérez-Gay et al., 2017; Pérez-Gay Juárez et al., 2019) we trained subjects to categorize visual textures with features that were global, holistic and hard to verbalize so as to minimize explicit, conscious categorization strategies. Comparing both similarity judgements and Event Related Potentials before and after categorization training, we found significant CP (between-category separation) accompanied by changes in an early negative component of the ERP (N1; 150-220 ms). The CP occurred only in those subjects who successfully learned the categories; those who failed to learn (given the same number of training trials) showed no CP effects, and the size of the effect was correlated with the size of the N1 component -which has been source localized to the extra-striate visual cortex and linked to processes such as feature extraction, supporting a true perceptual change after learning the categories. We also developed a deep learning neural net model of category learning with these textures composed of distributed, holistic features.

Despite the growing body of experiments showing perceptual changes after learning a category (Pérez-Gay Juárez et al., 2019; Livingston, Andrews, & Harnad, 1998; Maier & Abdel Rahman, 2019; Notman, Sowden, & Emre, 2005; Pérez-Gay Juárez et al., 2017); for a review, see (Andrews, De Leeuw, Larson, & Xu, 2017), the mechanisms behind these effects have not yet been fully explained. We have hypothesized that, if learning categories implies detecting the features that co-vary with membership, this process would induce a "feature-filter" -similar to those we have for innate categories. This learned filter would then automatize the categorization process, making relevant features "pop-out" instead of having to consciously look for them every time we encounter a stimulus.

Features can be thought of as dimensions in our psychological space. According to our previous computational models and empirical work, Learned CP effects occur as a product of Dimensional Reduction: By sorting the stimuli by trial and error, subjects learn to detect covariant features and ignore non-covariant features, reducing the dimensionality of perceived similarity space from N (total number of features) to k (number of features co-varying with category membership) (Pérez-Gay, Sicotte, Thériault, & Harnad, 2019; Thériault, Pérez-Gay, Rivas, & Harnad, 2018).  But dimensions (or features) are not always of the same nature. Some features we use to categorize stimuli are global, like texture, color or shape of an object, and some are localized, like object parts (Folstein, Palmeri, Van Gulick, & Gauthier, 2015). Previous research has shown differences in the time course of perceptual processing of global and local features that may impact object classification (Gerlach & Poirel, 2018; Navon, 1977).  To corroborate if the mechanisms for feature detection that make categories "pop-out" and their corresponding neural correlates are independent of the nature of the features, we now apply our previous methodology to learning to categorize computer-generated fish shapes with local and verbalizable features.

## 2. Materials and Methods

### 2.1 Participants

One hundred subjects (64 Females, 36 Males) aged 18 - 35 years were recruited online through the McGill Human Participant Pool. They were either native English-speakers or native French-speakers and free of significant neurological or psychiatric conditions. They read and signed the consent form approved by the *Comité institutionnel d'éthique de la recherche avec des êtres humaines*, Université du Québec à Montréal, approval number 803_e_2017.

### 2.2 Stimuli generation

We used computer-generated images of fish line-drawings with features resembling those used by Sigala & Logothetis, 2002. The fish had four binary features that could covary systematically with category membership (circular/semicircular side fin, rounded/straight upper fin, extended/foreshortened mouth and rounded/straight tail), plus two random features (clusters of dots that varied in shape and location, unrelated to category membership). There were two categories — KAILFISH and LIMFISH— and four possible proportions of category covariant features (k/N -4/6 to 1/6). Based on these rules, we generated samples of 160 fish images to be used as training sets for category learning (Figure 1); each image presented one or two times for a total of 200 trials. Not all the possible combinations of features were tested for each k/N level. Two sets of feature-combinations were randomly selected for use at level 4/6, 3/6, and 2/6 (for example, set 1 in level 2/6 would have the side fin and the upper fin as diagnostic features and set 2 would have the tail and the mouth as diagnostic features). For level 1/6, each of the four features was used as a diagnostic feature. Each subject was randomly assigned to a k/N proportion (1/6 - 4/6) and a feature-combination set.

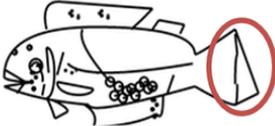

*Figure 1* **Sample of the stimuli** highlighting the possible values for each of the four binary features and showing examples of the variation of two non-binary features (dots location and shape). At the easiest level (4/6), all four binary features would have value 0 in Kailfish (K) and value 1 in Limfish (L). At the hardest level (1/6), only one of the four binary features would covary with category membership. The dot clusters varied in shape and location independently of category membership.

### 2.3 Procedure

The experiment was conducted in a sound isolated chamber with dim lighting and no other sources of electromagnetic interference. Subjects were seated in a comfortable armchair in front of a glass window through which they saw the computer screen presenting the task. They had a keyboard placed on a table between them and the window to click on the numerical keys, and the K and L keys. Sixty-four electrode channels were used to record whole-head EEG data through the Biosemi Actiview2 amplifier. The tasks were built and presented using the PsychoPy2 psychology open source software (Peirce, 2007).

## 2.4 Tasks

For this experiment, three tasks were administered: A standard reinforcement category learning task (trial and error with corrective feedback) preceded and followed by a pairwise dissimilarity judgement task.

The reinforcement learning task involved a process of trial and error, with participants attempting to categorize and receiving corrective feedback. Participants saw a total of two hundred fish (each stimulus appearing 1-2 times during the training), which they had to categorize as either a "KAILFISH" or a "LIMFISH".

Each trial consisted of a fixation cross (500 ms) followed by one of the stimuli, shown at the center of the screen against a white background (1 s). Subjects were instructed to click K or L to indicate the category. They had to respond within 2s of the onset of the stimulus; if they did not, the computer prompted them to respond faster. Responses were followed by immediate feedback (lasting 750 ms) indicating whether their response had been correct or incorrect. The inter-trial interval was 2500 ms. The 200 trials were divided into four blocks of fifty stimuli each. Following each block, there was a pause in which participants had to fill out a questionnaire asking whether they thought they had detected the difference between the KAILFISH and the LIMFISH. If they replied "yes", they were asked to describe what they were doing to categorize the stimuli. If they replied "no", they were asked to describe the provisional strategy they were using to try to sort them. The training session lasted about thirty minutes (pauses included).

In the dissimilarity judgements before and after the training, subjects rated eighty pairs of stimuli, presented simultaneously, for similarity. They were shown a fixation cross (500 ms), after which two stimuli appeared at the center of the screen for 750 ms each, one after the other, with an inter-stimulus interval of 1 s. Subjects were then asked to rate the dissimilarity on a scale of 1 to 9, such that 1 corresponded to "very similar" and 9 to "very different". They were encouraged to make use of the full range of the scale. Of the total of eighty pairs presented, 40 were within-category pairs (20 "Kailfish" and 20 "Limfish") and 40 were between-category pairs (but of course before training participants did knew neither the categories nor their names). Responses and reaction times were recorded during the task. The very same set of 80 stimulus pairs was presented in the same order for the dissimilarity judgements before and after training to all subjects at all the levels of difficulty. Following the first set of dissimilarity judgements, participants began their visual category training trials with corrective feedback, as mentioned above (200 trials divided into four blocks with questionnaires in each pause).

Instructions and questionnaires were in English or French depending on the subjects' native language. We recorded responses and reaction times during the three tasks (dissimilarity judgements and training).

## 2.5 EEG Acquisition

A Biosemi 64-electrode international reference cap was placed on the Ss' heads according to head circumference; electrodes were connected to the cap using a column of Conductive Gel to fill the gap between the skin and the electrodes. Six facial electrodes were placed at the common reference sites: two earlobes, above and below the right eye to record the VEOG (Vertical Electrooculogram), directly to the side of the left eye and directly to the side of the right eye to record the HEOG (Horizontal Electrooculogram). The signals were received by a Biosemi ActiveTwo amplifier at a sampling rate of 2048 Hz with a band pass of 0.01- 70 Hz. Impedance of all electrodes was kept below 5kOhms. Data collection was time-locked to time point zero at the onset of visual stimulus presentation.

**2.6 EEG Data Analysis**

EEGLab 13.4.4b open source software (Delorme and Makeig, 2004) was used to pre-process raw EEG files via the following steps: a) Down-sampling of the data to 500 Hz to decrease computational requirements; (b) Data cleaning using a low-pass (100 Hz) filter, high pass (3 Hz) filter and notch filter (60 Hz); (c) Identification of bad channels using EEGLAB function for chanel statistics (excluding data values below the 5 th percentile and above the $95^{th}$ percentile) (d) Interpolation of identified bad channels. (e) Re-referencing of electrodes to a virtual average reference including all head electrodes but excluding the facial ones. (f) Epoching the data into 3000 ms segments with individual epochs spanning from 1000 to 2000 ms around time zero. (g) Baseline correction using the 200 ms before each stimulus onset. (h) Identification of Independent Components using EEGLAB function *Runica* (Makeig, Jung, Bell, & Sejnowski, 1996) (i) Blink and eye-movement rejection after visually inspecting the topography and power spectrum of the first 10 components (j) Rejection of additional artifacts and noisy epochs using an extreme value filter (+/- 100 μV) and then a probability filter with a 2 standard deviation limit for single channel and a 6 standard deviation limit across channels. (k) Splitting the data into two parts—before learning and after learning for the Learners or first-half and second-half for other subjects.

After splitting datasets, grand averages were computed for comparisons within subjects (before vs. after learning or first half vs. last half trials) and between subjects (learners vs. non-learners). Event Related Potentials waveforms were plotted and quantified by extracting mean, minimal and maximal voltages in our windows of interest: the N1 and LPC -signatures of category learning and CP in previous studies (Pérez-Gay Juárez et al., 2019; Morrison, Reber, Bharani, & Paller, 2015) and P1 following other studies that have reported effects of categorization on perception (Forder, He, & Franklin, 2017; Maier & Abdel Rahman, 2019; Maier, Glage, Hohlfeld, & Abdel, 2014; Samaha, Boutonnet, Postle, & Lupyan, 2018). Scalp distributions were also plotted for each condition in the time-windows of interest. Statistical analyses on amplitude differences between conditions were assessed with repeated measures ANOVAs and post-hoc tests using the IBM SPSS 23 Statistical Software. Effect sizes were calculated using partial eta squared ($η^2$) and Cohen's d.

**2.7 Analysis of Dissimilarity Judgements**

We computed six variables from the dissimilarity judgment data. We averaged each subject's dissimilarity ratings for within-category and between-category pairs, pre and post training, creating the variables "within-category pre-training" (Wpre), "within-category post-training" (Wpost), "between-category pre-training" (Bpre) and "between-category post-training" (Bpost). We computed the variable difffW, for "within-category dissimilarity change" (Wpost – Wpre) and diffB for the "between-category dissimilarity change" (Bpost – Bpre). Finally, we computed the variable "Global CP" (diffB – diffW). The purpose of this last variable is to evaluate the overall CP effect and differentiate overall compression or separation from categorical perception. If diffW and diffB go in opposite directions, as they should in classic Learned CP effects (within-category compression and between-category separation), then substracting diffB – diffW will amplify Global CP. If diffW and diffB have similar values or go in the same direction, Global CP will be reduced.

The significance of diffB, diffW and Global CP was tested with repeated measures ANOVAs with "learning" as the between-group factor and t-tests to compare Successful Learners and Non-Learners separately. We also ran Pearson correlations for the Learners' similarity changes (within and between categories) and the significant changes in ERP components (peaks and amplitudes).

### 2.8 Neural net model of category learning of stimuli with local features

To complement our study of CP in *humans,* we used a preliminary method from upcoming work – aimed at identifying the computational and mathematical foundations of CP across the different machine learning models – to show the effects of category learning in artificial neural nets. We therefore trained a model using an autoencoder architecture, described in figure 2, with convolutional neural networks (CNNs) as opposed to the multi-layer perceptrons (MLPs) used in (Thériault et al. 2018). A technical overview of these methods is outside the scope of this article and will be covered in an upcoming publication, but see (Goodfellow, Bengio & Courville, 2016, pp. 326-366) for a general overview of CNNs and (Bronstein Bruna, Cohen & Velickovic, 2021, pp. 68-77) for a more mathematical one.

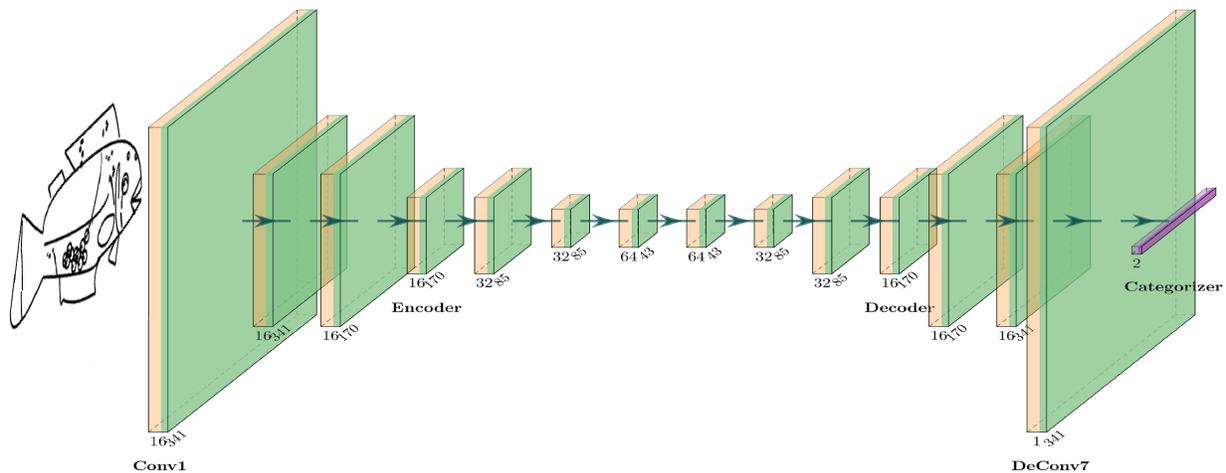

*Figure 2 :* **Our used autoencoder architecture**. It has 7 convolutional layers for the encoder and 7 deconvoluational layers for the decoder. A fully connected layer is added as a ''categorizer''. We used strided convolutions instead of max pooling as the latter decreases the precision of the decoding process. We deliberately overfitted the model on the data due to the small dataset but also to have better visualizations. Upcoming work will present results on larger datasets.

We trained our model to perform the task on the easiest level (4/6) on a limited dataset. This will make our results more of a proof-of-concept for using CNNs as computational models of CP. To study CP in our model, we implemented an algorithm to visualize the learning process via the highlighting of pixels that maximizes the activation of neurons in the last layer. It is openly inspired from (Mordvintsev, Olah, & Tyka 2015) and uses gradient ascent optimization. Our other method for studying CP in CNNs is the evolution in representational spaces of neurons following category learning via various distance metrics, in a spirit akin to *categorality* in (Bonnasse-Gahot, & Nadal 2022). The one presented here will be *diffW* and *diffB* described, in (Pérez-Gay et al. 2017). These are simple Euclidian distances of the different point in the representational space taken from all neurons at all epochs of the training.

# 3. Results

### 3.1 Learning assessment

Out of the hundred subjects we recruited, 9 subjects were excluded due to noisy EEG data or withdrawal before the completion of the task. Consequently, 91 subjects aged 18-36 (mean=21.89, 33 male, 57 female) completed our visual category-learning task, randomly assigned to one of four sets with different proportion of covarying features (k/N). Overall, 40 of the 91 subjects successfully met our a-priori learning criterion (reaching and maintaining at least 80% correct). Four of these subjects already had an accuracy of over 80% from the very outset of the categorization task, suggesting they had learned during the unsupervised phase (dissimilarity judgements) and were classed as "Immediate (or Unsupervised) Learners" and excluded from further analysis. The remaining subjects (51) did not reach the learning criterion and were classed as Non-Learners.

Repeated-measures ANOVA with Learning as a between-subject factor showed that the change in Reaction Times and Response Accuracy across the four successive blocks was significantly different between Learners and Non-Learners. (Accuracy: $F(3,92) = 40.197$, $p <0.001$, $\eta2 = 0.567$; reaction times: $F(3,92) = 7.651$, $p <0.001$, $\eta2 = 0.200$).

Subsequently, within-subjects one-way ANOVA confirmed that Learners' response accuracy increased linearly throughout blocks ($F(3,31)=69.459$, $p<0.01$, $\eta2=0.870$) while reaction times decreased linearly ($F(3,31)=21.835$, $p<0.01$, $\eta2=0.679$). Non-Learners showed changes across blocks on reaction times ($F(3,53)=8.561$, $p<0.01$, $\eta2=0.326$), but not on accuracy ($F(3,53)=1.936$, $p=0.135$, $\eta2=0.099$).

Based on our previously sketched Neural Net Model of category learning, we were interested to test if a lower the proportion of features covarying with category membership (k/N), made it harder to learn the category. This was not the case. A one-way ANOVA did not find any effects of the proportion of covarying features (k/N) in learning speed (number of trials to learning criterion, $F(3,33)=0.559$, $p=0.646$) or performance (percent correct in the last block) depending on k/N (number of trials: $F(3,33)=0.559$, $p=0.646$; accuracy: $F(3,33)=0.745$, $p=0.534$).

**3.3 Dissimilarity Judgements**

Subjects rated pairwise dissimilarity before and after the reinforcement learning task, which consisted of 200 trials with corrective feedback. Following the results of previous work (Pérez-Gay, Sicotte, Thériault & Harnad, 2018), we expected the Learners to rate between-category pairs as more different (separation) and within-category pairs as more similar (compression) after the training (the learned CP effect). To corroborate that these effects are a consequence of learning, they should be absent from subjects who undergo the same training but fail to reach the learning criterion (80%).

For our first assessment, we collapsed the data from the different sets and difficulties and ran a three-way repeated measures ANOVA to determine the effect of training on pairwise similarity for Learners and Non-Learners (we excluded the immediate learners from this analysis, to be able to fulfill the equality of variances assumption). There was a significant three-way interaction between learning (Learner vs. Non Learners), pair type (between-category vs. within-category) and training (pre-training vs. post-training), $F(1, 90) = 36.794$, $p < 0.001$, $\eta2=0.290$ (Figure 3). For the simple two-way interactions and simple main effects, a Bonferroni correction was applied, leading to statistical significance being accepted at the $p < .025$ level. There was a significant simple two-way interaction between pair type (between or within) and training (pre vs. post) for Learners, $F(1, 34) = 45.963$, $p < .001$, $\eta2=0.575$, but also for Non-Learners, $F(1, 56) = 7.257$, $p = 0.009$, $\eta2=0.115$, suggesting that training induces different effects in between-category pairs vs. within-category pairs change in both groups.

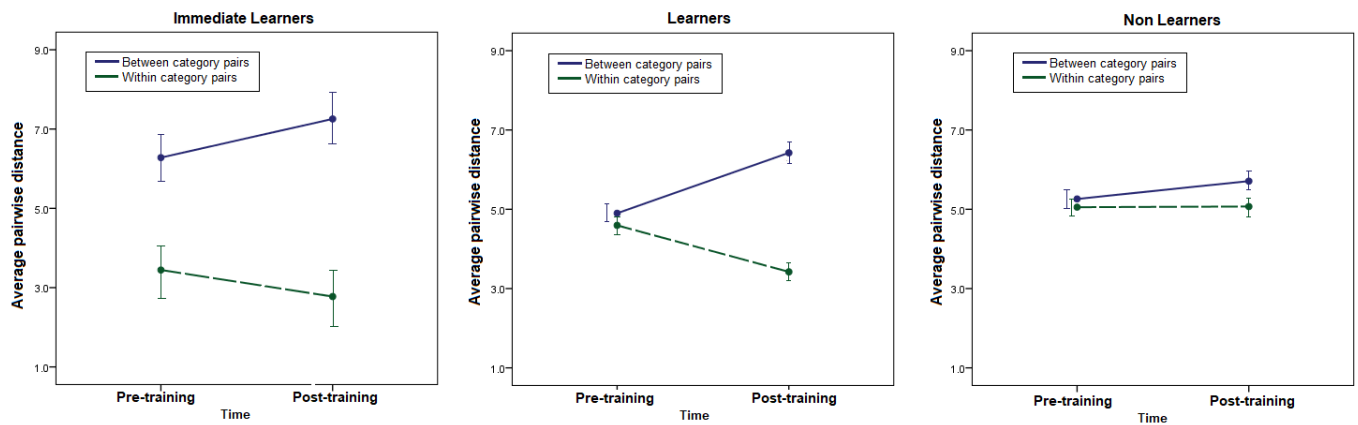

*Figure 2* **Between and within-category pairwise similarity judgements before and after the training for Immediate Learners (n=5), Learners (n=35) and Non-Learners (n=57).** *The pre-training scores of between and within category pairs (blue vs. green lines) were not significantly different between learners and Non-Learners, $t(90)=6.066$, $p<0.001$*

To test for between-category and within-category changes separately in Learners and Non-Learners, we ran a series of paired-samples t-tests. Learners rated between-category pairs as significantly "more different" after training compared to before (between-category separation; mean $diffB=1.5242$, $t(34)=5.879$, $p <0.001$, Cohen's $d= 1.004$) and within-category pairs as significantly less different after training compared to before (within-category compression; mean $diffW=-1.1721$, $t(56)=-3.705$, $p=0.001$, Cohen's $d= 0.631$). Non-Learners did not show significant changes in their dissimilarity scores for within-category pairs comparing before and after the training (mean $diffW=-0.0153$, $t(56)=-0.083$, $p=0.934$, Cohen's $d= 0.011$) but they did show a small, yet significant increase in the rating of between category pairs (mean diffB=-0.45, t(56)=-3.152, p=0.003, Cohen's d= 0.417).

As explained in the methods section above, we also computed a variable to assess compression and separation simultaneously: "Global CP" (diffB-diffW). Because between-category separation is expressed by a positive diffB and within category compression is expressed by a negative diffW, a subject showing both within category compression and between category separation (positive diffB and negative diffW) would have a higher value of Global CP. If a subject showed between-category separation but no within-category compression, then diffW would be near to zero or positive, resulting in a smaller value of Global CP. Using this variable, we ran a Mann-Whitney independent samples test to confirm that the "Global CP" values were significantly higher for Learners (mean=2.6964, SE=0.397) than for Non-Learners (mean=0.4346, SE=0.161), U=353.5, p<0.001.

### 3.4 Dissimilarity Judgements and Proportion of Covarying Features

We ran a three-way repeated measures ANOVA (rANOVA) to test for differences in the observed separation (diffB) and compression (diffW) effects for Learners assigned to different k/N levels. This analysis failed to show a significant interaction between training (pre-training vs. post-training), pair type (between-category vs. within-category) and number of covarying features ($k$/N), F(3,31)=0.224, p=0.865. $\eta2$=0.023. We also ran separate two-way rANOVAs that failed to show an interaction between training and $k$/N for separation (diffB, F(3,31)=1.168, p=0.338, $\eta2$=0.102) and also for compression (diffW, F(3,31)=1.143, p=0.347, $\eta2$=0.100), suggesting that, as was the case with the texture stimuli in our previous experiment (Pérez-Gay et al., 2019) the number of covarying features ($k$/N) did not have an effect on compression and separation.

We ran correlations between the number of trials to reach criterion (in Learners only) and the size of the CP variables. We found significant correlations for compression (rho=0.387, p=0.026), separation (rho=-0.350, p=0.046) and also Global CP (rho=-0.561, p=0.001), showing that the earlier the learning (fewer trials to criterion), the bigger the CP effects exhibited by learners, regardless of the k/N levels.

### 3.5 ERP results

Grand average ERPs were computed combining the data from the four difficulty levels, separating our subjects in Learners and Non-Learners. The Learners data were divided into trials before and after learning, while the Non Learners' were divided in half (first half versus second half). After applying the rejection methods described in Section 2.6, an average of 6.2 trials (5.6%) were rejected from the Learner's *before* trials (range: 0 to 25 trials, 0–18%) and about 5.38 trials, (5.38%) were rejected from the Learner's *after* trials (range: 0 to 24 trials, 0–21%) per subject. For Non-Learners, an average of 5.98 (5.98%) trials were rejected in the first-half (range: 0-21 trials, 0-21%) and an average of 7.17 (7.17%) were rejected from the second half (range: 0-22 trials, 0-22%). Two Non-Learners were excluded for having more than 25% rejected trials.

***3.5.1 Early P1 and N1 effects*** We computed within-subjects grand average ERPs for our two groups: Learners and Non-Learners, combining the four difficulty levels. We plotted the waveforms for the 64 electrodes for our first exploration of possible differences between conditions (before vs. after or first-half vs. second-half) to determine our ERP time-windows and Regions of Interest. This comparison revealed two significant changes in early ERP components when comparing trials before and after training: an increase in the occipital and parieto-occipital P1 positivity and a decrease in central N1 negativity. None of these effects was significant for the Non-Learners, when comparing the first half to the second half of the trials. We plotted scalp distributions to assess the topography of these effects in the P1 window (80-140 ms) and the N1 window (150-220 ms). To identify the electrodes that showed significant changes after learning, we used the paired samples t-test with FDR correction included in the EEGLab statistical package. Applying this correction for multiple comparisons, the central N1 effect in the Learners' lost its statistical significance. The scalp plots and ERP waveforms for the P1 effect can be observed in Figure 4.

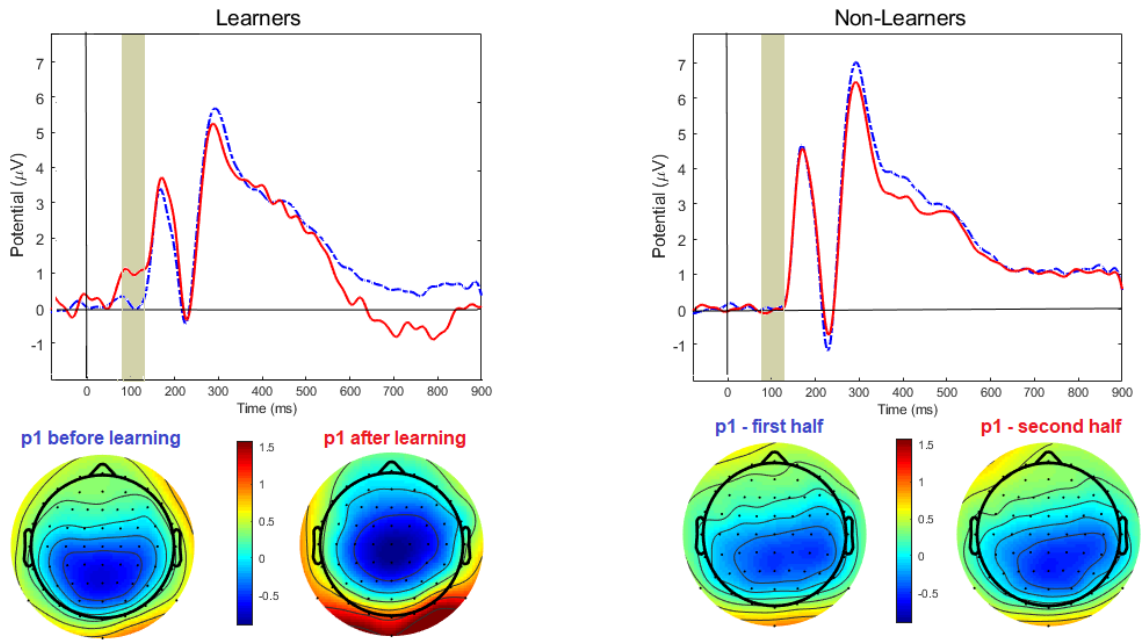

*Figure 4* **Event Related Potentials in Parieto-Occipital electrodes (averaging right- PO8 and left PO7). Above Left:** Within-subject ERP grand averages for Learners (n=35) across trials before vs. after reaching criterion for learning. **Above Right:** Within-subject ERP grand averages for Non-Learners (n=57), for the first vs. second half of trials. Highlighted in gray is the P1 window (80-140 ms). **Below**: Scalp maps comparing the before and after / first half and second half conditions in the P1 window (80-140 ms).

To perform additional statistical analysis on the P1 values, we extracted the amplitudes and peaks of this component in both Learners and Non-Learners. We chose electrodes Oz (midline occipital), PO7 (left parieto-occipital) and PO8 (right parieto-occipital), to assess for lateralization effects. We ran two-way repeated measures ANOVAs to determine the effect of training on the P1 amplitude of each electrode for Learners and Non-Learners. There was a significant two-way interaction between learning group (Learner vs. Non Learners), and training (before/first half vs. after/second-half) on the P1 amplitude at PO7, $F(1, 90)= 6.906$, $p=0.010$, $\eta2=0.071$; Oz, $F(1,90)= 17.880$, $\eta2=0.166$, and PO8, $F(1, 90)=14.241$, $p<0.001$,

η2=0.137. A two-way repeated measures ANOVA failed to show a significant interaction between learning (before and after) and hemi-scalp (right-PO8 vs left-PO7), $F(1,34)=2.407$, $p=0.130$, η2=0.066), indicating that this effect was not lateralized; no significant differences between the right and left P1 increase throughout the category learning task were observed in the Learner's group. We also ran a last two-way repeated measures ANOVA that failed to show a significant interaction between learning (before vs. after) and difficulty on the Learner's P1 amplitude increase at PO7, $F(3,31)= 1.037$, $p=0.390$, η2=0.091; Oz, $F(3,31)= 0.960$, η2=0.085, and PO8, $F(3, 31)=1.528$, $p=0.227$, η2=0.129

Finally, a series of t-tests confirmed that, for the Learners, the P1 amplitudes and peaks were significantly bigger for the trials after training compared to before in the three occipital electrodes (PO7, Oz, PO8), being maximal at P08. In contrast, Non-Learners did not show any significant changes in these parameters between the first and second half of their trials except for a decrease in the P1 amplitude in electrode Oz that disappeared when correcting for multiple comparisons (level $p<0.025$) (Table 2).

*Table 2*
*Paired samples t-tests for P1 peaks and amplitude changes throughout the training.*

|  | Learners (before-after) | | | | | Non-Learners (1st half – 2nd half) | | | | |
| --- | --- | --- | --- | --- | --- | --- | --- | --- | --- | --- |
|  | *Mean* | *T* | *Df* | *p* | *d* | *Mean* | *t* | *df* | *p* | *d* |
| PO7 - peak | -0.7099 | -2.469 | 34 | **0.019** | 0.438 | -0.0241 | -0.1616 | 56 | 0.932 | 0.011 |
| PO7 - amplitude | -0.7344 | -3.000 | 34 | **0.005** | 0.554 | -0.0155 | -0.085 | 56 | 0.872 | 0.021 |
| Oz - peak | -0.7832 | -3.247 | 34 | **0.003** | 0.606 | 0.0234 | 1.571 | 56 | 0.122 | 0.210 |
| Oz - amplitude | -0.6006 | -3.065 | 34 | **0.004** | 0.555 | 0.2672 | 1.991 | 56 | 0.051 | 0.264 |
| PO8 - peak | -1.0929 | -3.473 | 34 | **0.001** | 0.604 | 0.1573 | 0.835 | 56 | 0.407 | 0.111 |
| PO8 - amplitude | -1.0568 | -4.202 | 34 | **<0.001** | 0.806 | 0.1676 | 1.008 | 56 | 0.318 | 0.133 |

### 3.5.2 Late ERP effects

In Learners, we found an increase in the parietal positivity (LPC- late positive component) between 400-600 ms when comparing the trials before reaching the learning criterion to those after reaching it. This difference was significant in a cluster of parietal electrodes (P1, Pz, P2, CP1, CPz, Cp2, P3, P4) and maximal at Pz. In contrast, the Non-Learners, showed a reverse pattern: a small decrease in this late positivity that was not consistent across the chosen electrodes (Figure 5).

We extracted the means and amplitudes of the LPC in the parietal cluster and three individual parietal electrodes (Pz, CP3 (left) and CP4 (right)). We ran two-way repeated measures ANOVAS to determine the effect of training on the LPC amplitude of each electrode for Learners and Non-Learners. We found a significant two-way interaction between learning group (Learners vs. Non Learners), and training (before/first half vs. after/second-half) on the LPC amplitude in the chosen cluster, $F(1, 90)= 40.306$, $p<0.001$, $\eta 2=0.309$. A two-way repeated measures ANOVA failed to show a significant interaction between learning (before and after) and hemi-scalp (right-CP4 vs left-CP3) on the LPC amplitude, $F(1,34)=1.931$, $p=0.174$, $\eta 2=0.054$), indicating that this effect was not lateralized; no significant differences between the right and left LPC increase throughout the category learning task were observed within the Learner's group. To test for potential differences between difficulties, we ran a last 2x2 ANOVA that failed to show a significant interaction between learning (before and after) and difficulty level (4/6, 3/6, 2/6, 1/6) on the LPC amplitude, $F(3,31)=1.789$, $p=0.170$, $\eta 2=0.148$.

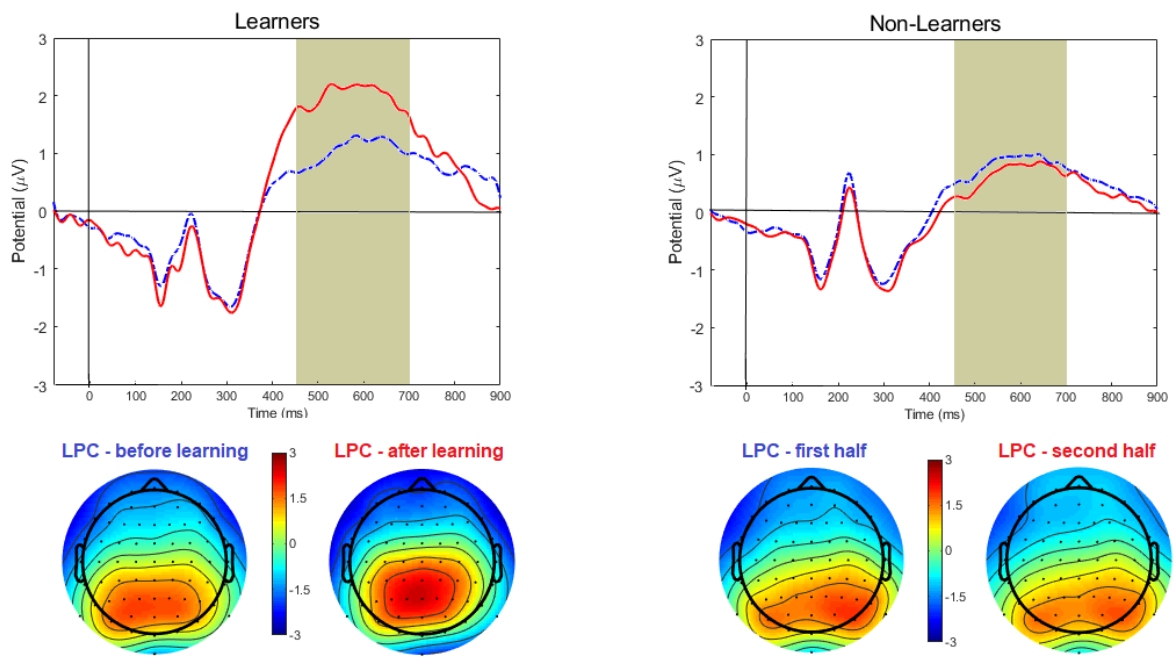

*Figure 5* **Event Related Potentials in a cluster of Parietal electrodes (Cz, CPz, CP1, CP2). Above Left:** *Within-subject ERP grand averages for Learners (n=35) across trials before (blue) vs. after (red) reaching learning criterion.* **Above Right:** *Within-subject ERP grand averages for Non-Learners (n=57), for the first (blue) vs. second half (red) of trials. Highlighted in gray is the LPC time-window.* **Below**: *Scalp maps comparing the before and after conditions in the LPC window (450-700 ms).*

Finally, a series of t-tests confirmed that, for the Learners, the LPC amplitudes and peaks were significantly bigger after learning trials compared to before in the chosen parietal cluster and three additional parietal electrodes (PO7, Pz, PO8), being maximal at P08. In contrast, Non-Learners did not show this consistent increase in both peaks and amplitudes. Some electrodes show a change in the opposite direction: a small but significant decrease in peak and amplitude of the LPC between the first and the second half of the trials (Table 3).

*Table 3.*
*Paired samples t-tests for LPC peaks and amplitude changes throughout the training.*

|  | **Learners (before - after)** | | | | | **Non-Learners (1st half – 2nd half)** | | | | |
|---|---|---|---|---|---|---|---|---|---|---|
|  | *Mean* | *T* | *Df* | *p* | *d* | *Mean* | *t* | *df* | *p* | *d* |
| Cluster - peak | -1.3871 | -5.662 | 34 | **<0.001** | **1.013** | 0.24785 | 1.715 | 56 | 0.092 | 0.229 |
| Cluster - amplitude | -1.1427 | -4.916 | 34 | **<0.001** | **0.913** | 0.40655 | 2.936 | 56 | **0.005** | **0.39** |
| CP3 - peak | -0.9556 | -5.076 | 34 | **<0.001** | **0.916** | 0.29702 | 2.261 | 56 | **0.028** | **0.3** |
| CP3 - amplitude | -0.7344 | -4.652 | 34 | **<0.001** | **0.852** | 0.40577 | 3.375 | 56 | **0.001** | **0.447** |
| Pz - peak | -1.2970 | -4.440 | 34 | **<0.001** | **0.772** | 0.33710 | 1.929 | 56 | 0.059 | 0.257 |
| Pz - amplitude | -1.1288 | -4.519 | 34 | **<0.001** | **0.805** | 0.48565 | 2.837 | 56 | **0.006** | **0.279** |
| CP4 - peak | -1.5614 | -6.503 | 34 | **<0.001** | **1.134** | 0.34912 | 2.751 | 56 | **0.008** | **0.366** |
| CP4 - amplitude | -1.2101 | -5.126 | 34 | **<0.001** | **0.93** | 0.17521 | 1.272 | 56 | 0.208 | 0.217 |

### 3.6 Correlations between behavioural and ERP effects

To better understand the relationship between the behavioral and the physiological variables, we computed a series of Spearman rank-order correlations between our ERP parameters (P1 and LPC amplitudes) and our behavioral changes (Reaction Times, response Accuracy and perceived similarity). Previous studies (Pérez-Gay, Sicotte, Thériault, & Harnad, 2018) found that early ERP components were correlated with perceptual effects (a positive correlation between N1 and the separation effect –diffB) while later ERP components correlated instead with learning indices (accuracy, reaction times). Both Learners and Non-Learners were included in the analysis.

We computed the correlations between our learning indices (Accuracy and Reaction Times in the last block) and our ERP amplitudes (P1 before/1st half, LPC before/1st half, P1 after/2nd half, LPC after/2nd half, P1 change, LPC change) to determine whether the behavioral changes were reflected in the physiological measures. We also computed the correlations between our CP measures (diffB, diffW and Global CP) and the amplitudes of our ERP components of interest (P1 and LPC) to determine whether the perceptual changes were reflected in the physiological measures.

### *3.5.4 ERP and CP effects*

Testing for DiffB and DiffW separately, we observed a significant negative correlation between the diffW and both the P1 after (rho=-0.214, p<0.05) and the P1 change (rho=-0.348, p<0.01): the more negative the diffW values (more compression), the bigger the P1 after learning and the bigger the P1 change. This correlation was significant only for the PO8 electrode. We also observed a significant, positive correlation between diffB and the P1 after learning in the PO7 electrode only (rho=0.236, p<0.05). No correlations were found between the P1 values before learning and diffB nor diffW.

Using the variable Global CP (DiffB – DiffW), we found more consistent correlations across our three electrodes. There were positive correlations between Global CP and P1 after learning in electrodes PO8 (rho=0.315, p=0.002), Oz (rho=0.262, p=0.012) and PO7 (rho=0.244, p=0.019) and also between Global CP and P1 change in electrodes PO8 (rho=0.39, p<0.001) and Oz (rho=0.212, p=0.042). Averaging the three electrodes together, Global CP is positively correlated with P1 after learning (rho=0.298, p=0.004) as well as with P1 change (rho=0.280, p=0.007). This shows that subjects with greater P1 amplitudes after learning and a greater P1 increase throughout the task exhibited a greater learned CP effect (Figure 6).

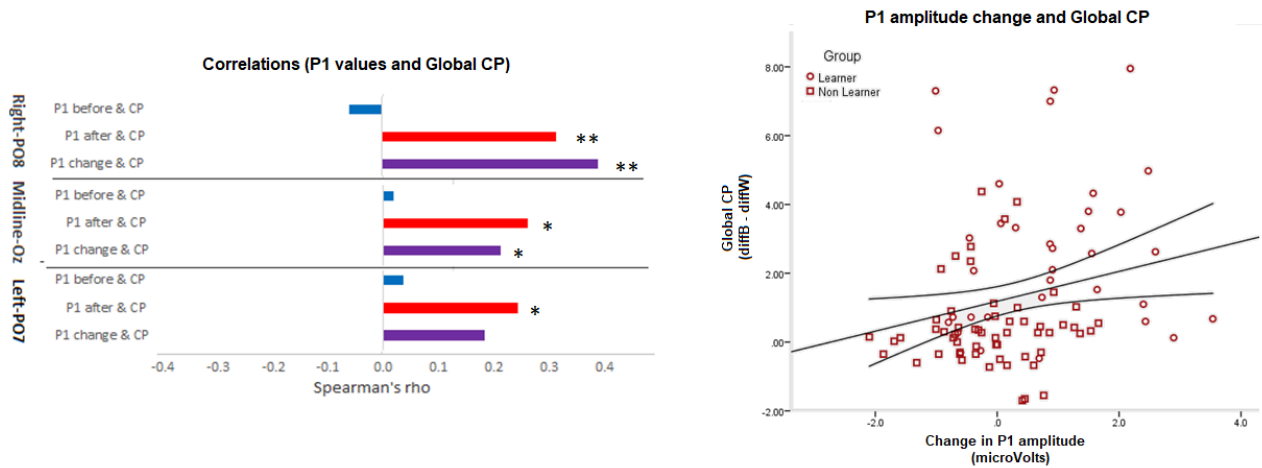

*Figure 6*. **Left**: *Spearman's rank-order correlations show that learned CP effects (Global CP: DiffB - diffW) are correlated with the P1 component amplitude (before and after learning, as well as with the before/after change). Significant correlations are indicated by asterisks (two asterisks if <0.01, one if <0.05).* **Right:** *Scatter plot for the correlation between P1 amplitude change (average of the three selected electrodes) and Global CP.*

Interestingly, in this experiment, the correlations between Global CP and ERPs were not restricted to early effects. We also found positive Spearman rank-order correlations between the LPC amplitudes in the parietal cluster and the Global CP variable for both the LPC after (rho=0.322, p=0.002) and the LPC change throughout the training (rho=0.447, p<0.001). This makes sense considering that, in our experiment, the P1 change was positively correlated with both LPC after learning (rho:0.251, p=0.016) and the before/after change in LPC (rho=0.390, p<0.001). To confirm that these effects are due to learning, the LPC before learning did not correlate with CP values.

### 3.5.5 ERP and Learning effects

As predicted, we found significant correlations between our learning indices and our LPC amplitudes. The Accuracy in the last block was positively correlated with the LPC after learning (rho=0.366, p<0.001) and the LPC change from before to after (rho=0.464, p<0.001), but not with the LPC before. The Reaction Times in the last block were negatively correlated with the LPC after learning (-0.214, p=0.022) and with the before/after change in the LPC (-0.266, p=0.011). This confirmed that bigger LPC increases corresponded to better performance in the categorization task (higher accuracy and faster reaction times) (Figure 7).

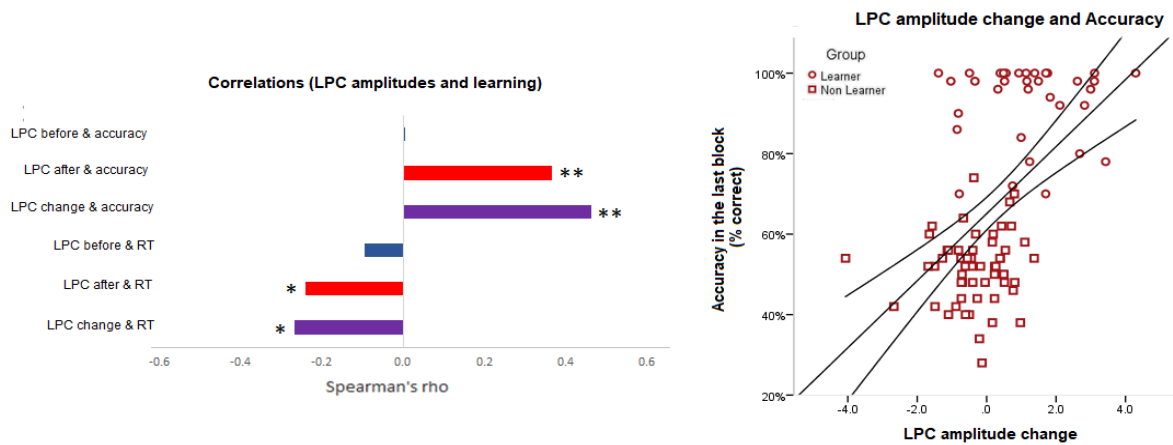

*Figure 7* **Left**: Spearman's rank-order correlations between learning indices (Accuracy and Reaction Times in the 4$^{th}$ block) and the LPC component amplitude (before, after and before/after change). Significant correlations are indicated by asterisks (two asterisks if <0.01, one if <0.05). **Right:** Scatter plot for the correlation between the LPC amplitude change (in our selected parietal cluster) and task accuracy (% of correct responses).

Analogously with what was found in the correlations between Global CP and ERPs, our correlations between ERP components and learning indices were not restricted to early effects. We also found positive Spearman rank-order correlations between the P1 amplitudes averaging our three electrodes of interest and the accuracy in the task's last block for both the P1 after learning (rho=0.332, p=0.001) and the P1 change during the training (rho=0.314, p=0.003). Reaction times were only correlated with the P1 change (rho=-0.237, p=0.025). As expected, neither the LPC nor the P1 before learning correlated with Accuracy. Two simple linear regressions were calculated to predict "Global CP" based on the P1 change and the LPC change respectively. A significant regression equation was found for the P1 ($F(1,90)=5.239$, p=0.024), with an R2 of 0.055. However, we also found a significant regression equation for the LPC change ($F(1,90)=5.239$, p<0.001) with a higher R2 value of 0.134. Participant's global CP increased 0.434 similarity rating units for every microvolt of P1 change and 0.548 similarity rating units for each microvolt of LPC change. These results indicate that both components can predict the overall CP. However, these results must be interpreted with caution because the P1 change and the LPC change are positively correlated. A potential explanation for the association between the early and the late components in this category learning task is elaborated in the discussion.

### 3.7 Results of the Neural Net model

Like in our prior work, the neural net model exhibited the expected separation and compression effects. The *diffW* and *diffB* exploded as soon as the supervised learning phase started and no further significant variation in the representational space of neurons occurred after. This suggests that once the local features are learned – near instantly in this case – the effects of CP occur in an all-or-nothing fashion. This contrasts with the more progressive evolution of CP in prior models that were trained with distributed, holistic features.

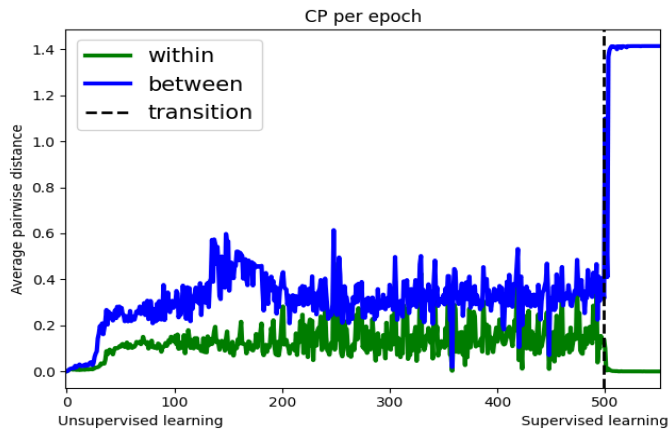

*Figure 9 :* **Evolution of CP through epochs**. This figure shows the initial separation and compression from mere exposure (unsupervised learning) as opposed to the explosion of CP when categorization occurs (supervised learning in this case). This effect is highly accentuated due to overfitting and small size of the dataset.

The other method used to evaluate CP through dimension reduction was the earlier-mentioned algorithm to identify which pixels maximized the activation for each neuron of each layer of our model. We report here only the data for a single neuron chosen on the last layer, using qualitative evaluation of the *filters visualizations*. This qualitative evaluation across all neurons of the last layer as shown that almost every single one of them has learned to recognize either both or one of these two features : the dorsal fin or the lower-end of the tail. The other two relevant features did not maximize the activity of any single neuron. This dimension reduction process over 30 supervised epochs is visualized in figure 10.

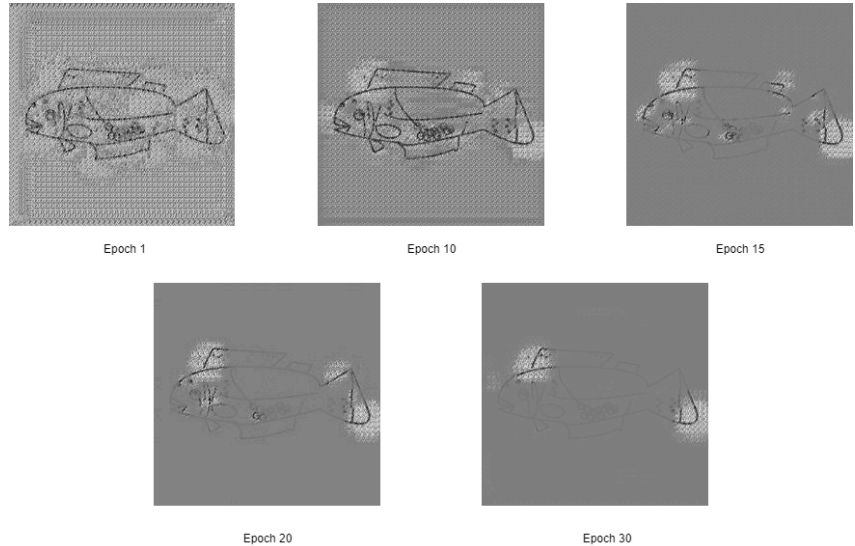

*Figure 10 :* **Evolution of Feature Learning.** This figure highlights the process of feature learning in a *toyish* CNN. In the beginning, the selected neuron has the same *activation* form almost all neurons but as the learning progresses, we can it the activation progressively narrowing down around the two features used to categorize : the dorsal fin and the lower-end of the tail.

## 4. Discussion

Building on previous work (Pérez-Gay, Sicotte, Thériault, & Harnad, 2018; Pérez-Gay et al., 2017), the aim of this experiment was to induce CP by training subjects to sort visual stimuli into two categories after one hour of trial and error training with corrective feedback. To test whether CP effects (between-category separation and/or within-category compression) were generated, we measured changes in behavior (accuracy and dissimilarity judgments) and electrophysiology (Event Related Potentials; ERPs). In the present experiments, in place of textures with globally distributed, micro-features we tested line-drawings of fish with local macro-features (shapes of parts), that could be detected and verbalized explicitly. The results shed more light on the role of spatial attention and explicit learning in visual category learning and resulting CP effects.

Before learning to categorize, subjects rated pairs of stimuli for degree of dissimilarity. The fish stimuli varied in many local features (mouth, side fin, upper fin, tail at fixed positions and clusters of dots that varied in location). To categorize correctly, subjects needed to detect the features that distinguished the members of one category (Kailfish) from the members of the other category (Limfish), i.e., the features that *covaried* with with category membership. Following 200 trials with corrective feedback, our successful learners (only) rated between-category pairs as significantly more different (separation) and within-category pairs as significantly more similar (compression). Subjects who had been trained with the same number of feedback trials but who had failed to meet our learning criterion (80%) had much smaller between-category separation and no within-category compression, confirming that the change in perceived dissimilarity depends mainly on learning the category, and only minimally on repeated exposure to the stimuli.

Successfully learning to categorize the fish correctly was accompanied by increases in the amplitude of two main ERP components, the occipital P1 (80-140 ms) and the parietal LPC (400- 600 ms). These effects were absent in subjects who failed to learn the category. The amplitude of both components also correlated positively with the size of the "Global CP" effect (combining separation and compression) as well as with accuracy across all subjects. The time-course and functional significance of these two components shed some light on the neural and cognitive processes underlying our learned CP effects.

### 4.1 Dimensional Reduction

Our candidate explanation for learned CP, supported by a neural network model (Thériault, Pérez-Gay, Sicotte, Rivas & Harnad, 2018) is *dimensional reduction*. In this model, the N features of the stimuli are treated as dimensions in an N-dimensional space. To categorize correctly, subjects must learn to selectively detect the k features that covary with category membership and to selectively ignore the N-k features that are irrelevant to category membership. Whereas the distance between pairs of stimuli in the hidden layer is based on N dimensions before learning the categories, it is reduced to the distances on the k covariant dimensions after successful learning. This is what generates the between-category separation and within category compression, making the category "pop out."

The fish stimuli we used in this experiment varied in six features (N = 6). Four of them -- the mouth, the upper fin, the side fin and the tail -- were binary and potentially covariant with category membership, with the number of covariants (k) being from 1 to 4, depending on experimental condition. The two other features that varied were the location and shape of the dot clusters; they were irrelevant in all cases. It is important to point out that even if all four features covaried, as in the easiest difficulty, it would have been enough to detect just one of them in order to categorize successfully. This is true whenever the covariant features are *conjunctive* (i.e, they all covary jointly) and local, i.e., separate from each other in space. Hence although finding at least one covariant may be easier and faster when there are more covariants, dimensions would be reduced from N to 1 rather than from N to k if only one were detected and used. Indeed, analysis of the subjects' categorization strategies showed that all reported using only one feature (though not necessarily the same one), irrespective of the value of k. As a consequence, the degree of dimensional reduction (and thus the size of CP) for all levels of k would be the same. This is indeed what we found: a strong CP separation/compression effect for all learners, irrespective of k, confirming the dimension-reduction effect, but with a reduction from N to 1. It will require more complicated modelling, as well as more complicated experiments (testing *inclusive disjunction* [either feature f or g or h or fg or fh or gh or fgh] as well as *exclusive disjunction* [f or g or h but not fg or fh or gh or fgh] to force the model and the subjects to use all k features rather than just one].

In human subjects, another difficulty in testing the dimensional reduction hypothesis fine-tuned to test k is feature inequality. When analyzing the number of trials needed to reach the learning criterion, we found that it did not depend on k, but more on which feature the subject picked as a strategy, confirming that some features are more salient than others to our visual system. This was already apparent in the texture experiments with distributed microfeatures but it becomes even more prominent when features have precise locations in space and gaze fixation plays a role in detecting them (Borji, 2012; Braun & Julesz, 1998; Itti, Koch, & Niebur, 1998) In the fish experiments with local macro-features, the side fin (located near the center of the fish) was the most used strategy among the Learners (even though the covarying features were counterbalanced across conditions). Because the resulting CP is correlated with the number of trials to learn (i.e., bigger CP effects for early learning), subjects show bigger CP effects for the most salient feature (the side fin), regardless of the number of covarying features (k). To test the model as an unbiased function of k would require either (1) an extremely large number of experimental condition as well as subjects to try to counterbalance for feature inequalities or (2) an attempt to simplify the features to make them more equal. This is all much easier to do in our computer model than with real visual stimuli and human subjects.

### 4.2 Learned CP and early ERP changes

Our ERP analysis revealed an increase in the occipital P1 -- the first positive ongoing component of the ERP, peaking around 100 ms -- after learning a new category. P1 is generated at the dorsal extrastriate cortex of the middle occipital gyrus (Di Russo, Martínez, Sereno, Pitzalis, & Hillyard, 2002; Di Russo et al., 2012). It is a part of the visual evoked potential thought to reflect the processing of low-level visual object features (Hillyard & Anllo-Vento, 1998). P1 is modulated by attention, linguistic cues and previous knowledge about the stimuli: bigger amplitudes can be seen with stimuli at locations to which we are paying attention (Baumgartner et al., 2018; Hillyard & Anllo-Vento, 1998; Kelly, Gomez-Ramirez, & Foxe, 2008) and with stimuli about which we have prior knowledge or that are "meaningful" to us (Lupyan, 2012; Maier, Glage, Hohlfeld, & Abdel, 2014; Samaha, Boutonnet, Postle, & Lupyan, 2018).

In our fish experiments, the P1 of the learners increased in amplitude after reaching the learning criterion, whereas those subjects who failed to learn showed no changes in this component throughout the task. This change in the P1 (80-140 ms) occurs slightly earlier than the effect we observed for textures, with their distributed features (Pérez-Gay et al., 2018); there the change was in N1 (150-220 ms), likewise an early perceptual component(Di Russo et al., 2002; Heinze et al., 1990; Slagter, Prinssen, Reteig, & Mazaheri, 2016; Vogel & Luck, 2000). The fish features are local, appear at a fixed spatial location and do not interact with each other. These properties would favor changes in the occipital P1, which is known to be modulated by single visual features appearing in cued spatial locations (Baumgartner et al., 2018; Heinze et al., 1990; Hopf et al., 2002), rather than the occipital N1, which is involved in discrimination and more holistic processing of visual stimuli (Fort, Besle, Giard, & Pernier, 2005; Morrison, Reber, Bharani, & Paller, 2015; Vogel & Luck, 2000). The fact that the fish features are easily nameable and already familiar (as fish organs) would also favor a P1 effect, in which prior knowledge of the stimuli would accelerate visual discrimination (Samaha et al., 2018), increasing the earlier P1 positivity instead of decreasing the later-occurring N1 negativity, which is modulated when learning to categorize unfamiliar textures (Pérez-Gay et al., 2018). As described in the results section, the P1 increase correlated positively with accuracy as well as with the size of the CP effect (separation and compression). This suggests that both the increased CP effect and the enhanced categorization performance may be associated with changes in the early sensory processing of our stimuli. Learned CP effects can thus be understood as the result of top-down modulation of early low-level perceptual processes (happening in modality-specific cortices) by learned categorical judgements stored in semantic memory (located in multimodal, association cortices).

### 4.3 Learned CP and late ERP changes

Besides the changes in the occipital P1, learning a category also induced an increase in the parietal LPC (late positive component). The LPC is a large, late parietal positivity of the P3 family. Its onset can vary between 400 and 600 ms and it peaks between 500 and 700 ms. Significant increases in LPC positivity for correct categorization trials as well as after learning to categorize compared to before have previously been reported in the literature (Bharani et al., 2016; Morrison et al., 2015; Pérez-Gay et al., 2018). These changes reflect decisional and other higher-order processes that take place during categorization, such as recollection of the categorization strategy during explicit learning (Paller, Voss, & Westerberg, 2009) or the link between perceptual information stored in working memory and potential responses during implicit learning (Verleger, Jaśkowski, & Wascher, 2005). The LPC increase with the fish stimuli (significant between 400-600 ms) occurred earlier than what we had observed with the textures (600-800 ms). This difference in latency indicates that the fish stimuli require less time to be evaluated and classified than the textures, probably because fish features are familiar, local and nameable. Overall, our LPC differences were robust both within subjects (after learning – before) and between subjects (comparing the LPC after learning to the LPC during the second half of the task) and the amplitude change was strongly correlated with the categorization accuracy, confirming that LPC is a reliable index of category learning.

**4.4 Perceptual effects, explicit learning and the cognitive bias**

A frequent objection to studies that report learned CP on the basis of similarity judgements is that the separation and compression effects observed could be just a naming bias from having learned the category name (rating stimuli as more different when their names are different and more similar when their names are the same) rather from than a perceptual change (Hanley & Roberson, 2011; Kikutani, Roberson, & Hanley, 2008; Pilling, Wiggett, Özgen, & Davies, 2003). Others in our laboratory have replicated the significant CP effect using objective psychophysical measures of discriminability (ABX or Same-Different judgements) and signal detection analysis in place of similarity judgments (Véronneau, et. al., 2017). Another factor making naming bias unlikely is that the categorization responses during the 200 training trials for both the textures and the fish were not naming at all, but motor responses (key-pressing K or L). In addition, the fact that in our texture experiments the CP effect was correlated with an early perceptual component of the ERP, N1, rather than with the later, more cognitive component, LPC, makes it still less likely that what underlies the CP is a naming bias.

Yet an important difference between the texture experiments and the fish experiments is that the texture features were distributed and hard to localize, whereas the fish features were readily and explicitly localizable. Following an explicit verbal rule based on a fixed local feature does make a bias possible: not a naming bias, but perhaps an attentional strategy bias. Once the subjects had detected one covariant feature, all they had to do was attend to that feature's fixed locus to know the category of fish. This is supported by the differences between the ERP findings for the textures and the fish: With the textures there was a dissociation between (1) the change in the early perceptual component (N1) of the ERP after learning, which was correlated with the size of the CP effect, and (2) the change in the later cognitive component (LPC), which was correlated with categorization performance (percent correct). With fish, in contrast, both the P1 change and LPC change after learning were correlated with the perceived change in similarity. This suggests that for explicit category learning of stimuli with local, easy to verbalize features, higher-order processes indexed by the LPC may also play a role in making the dissimilarity judgements.

Finally, it is important to bear in mind that reinforcement learning during a one-hour, 200-trial training session does not resemble category learning in real life and does not seem enough to induce durable perceptual changes. To assess "true" perceptual changes, longer-term studies with extended trials spaced across weeks are being conducted as a next step in this research.

**4.5 Category Learning, from Learned CP to linguistic relativity**
In some cases, differences between members and non-members of a category are obvious enough to pick up these features by passive exposure, just by looking at (or listening to) the exemplars without any feedback or instruction (e.g. telling apart elephants from zebras). This corresponds to *unsupervised learning* (Fisher, Pazzani, & Langley, 1991). However, in most other instances, the raw sensory input will not be enough to distinguish members and non-members (for example, in the case of edible vs. inedible mushrooms). In that case, to learn the category we would need to sort by trial and error using different, random features of the mushroom (color, size, texture, taste) and interacting with them until we manage to detect the features that covary with (and are thus predictive of) category membership. This corresponds to a supervised, *sensorimotor* category learning process (Cangelosi, 2017; LeCun, Bengio, & Hinton, 2015; Smith & Rangarajan, 2016).

Having to sort every instance of the world by trial and error would be an exhausting, time-consuming endeavor. Thanks to the development of language, humans have managed to learn and teach categories by *verbal instruction*, this is, after we learn the features that are relevant for categorization, we can use words to share them with others (e.g.: "mushrooms with red on the cap or stem are poisonous"), saving them the time and effort they would have needed it to discover it themselves.

Whether it is through trial and error or verbal instruction, learning to categorize will result in filtering the raw sensory input to detect reliable correlates – sensorimotor features – that reliably predict category membership. In previous papers, we have argued that evidence of Learned CP is also evidence supporting the "weak" version of the "Whorf-Sapir hypothesis", also referred to as linguistic relativity (Pérez-Gay et al., 2019; Pérez-Gay et al., 2017). This hypothesis suggests that the language we speak shapes our perception of the world (Boroditsky, 2006; Kay & Kempton, 1984; Regier & Kay, 2009; Zhong, Li, Li, Xu, & Mo, 2014). We add that it is not just "language", but *learning categories* what modifies our perception.

In most cases, human categorization entails assigning names, or linguistic labels, to categorized objects or situations (Lupyan, 2012). Most of the "content" words in natural languages -- nouns, verbs, adjectives and adverbs –are the names of learned categories (Bloom, 2000; Horst & Simmering, 2015). As pointed above, language also allows us to learn categories by combining words into definitions and descriptions, as in a dictionary. But to learn a new category from a definition, we have to already *understand* the words in the definition: The defining words already have to be "grounded" -connected to their referent in the world. Therefore, to be able to acquire our whole vocabulary, at least some words have to be grounded directly through sensorimotor category learning (Harnad, 1990; Vincent-Lamarre et al., 2016). If this analysis is right, it suggests that the mechanism underlying Learned CP – the learned feature filter that makes the category to which something belongs "pop out" perceptually – may be the mechanism that grounds the meanings of our words. Along the lines hypothesized by Whorf and Sapir, words would come to mean what they mean because we have learned to sort and name things. Experimental evidence for changes in perception after category learning of perceptual changes after learning a category would then support an influence of language on perception.

## 5. Conclusion

Our results provide further evidence for learned CP (both between-category compression and within-category separation), this time for stimuli with local, easy to verbalize features. Previous studies using texture stimuli with distributed rather than local features showed that category learning was accompanied by changes in early and late ERP components. This study confirms our previous findings: We report an increase in the parietal late positive component, involved in decisional stages of categorization as well as a change in an early, occipital component, which provides further evidence for changes in early stages of visual perception occurring as part of the category learning process. While the increase in the LPC component corresponds to what we had previously reported for texture stimuli, the early effect is observed in the P1 (80-140 ms) instead of the N1. Furthermore, the changes in early ERP components (namely the P1) were correlated with the size of the CP effect. We interpret the results of this study as further behavioral and physiological evidence of Learned CP effects, Further studies with longer, spaced training sessions are needed to corroborate these results.